\documentclass[aps,prl,twocolumn,showpacs,10pt,superscriptaddress,preprintnumbers,nofootinbib]{revtex4-1}
\usepackage{graphicx}
\usepackage{amsmath,bm,braket}
\usepackage{color}

\allowdisplaybreaks

\newcommand{\dd}{\mathrm{d}}

\newcommand{\als}{\alpha_s}

\newcommand{\nn}{\nonumber}

\newcommand{\orda}{\mathcal{O}(\alpha_s)}
\newcommand{\ordb}{\mathcal{O}(\alpha^2_s)}

\def\ord#1{{\mathcal{O}(#1)}}

\begin{document}

\preprint{SLAC-PUB-16101, SMU-HEP-14-08}

\title{Top-quark forward-backward asymmetry in $e^+e^-$ annihilation at NNLO in QCD}
\author{Jun Gao}
\email{jgao@anl.gov}
\affiliation{Department of Physics, Southern Methodist University, Dallas,
TX 75275-0175, USA}
\affiliation{High Energy Physics Division, Argonne National Laboratory, Argonne, IL 60439, USA}
\author{Hua Xing Zhu}
\email{hxzhu@slac.stanford.edu}
\affiliation{SLAC National Accelerator Laboratory, Stanford University,
Stanford, CA 94309, USA}

\begin{abstract}
\noindent
We report on a complete calculation of electroweak
production of top quark pairs in $e^+e^-$ annihilation at
next-to-next-to-leading order in Quantum Chromodynamics.
Our setup is fully differential and can be used to calculate
any infrared-safe observable. Especially we calculated the
next-to-next-to-leading order corrections to top-quark
forward-backward asymmetry and found sizable effects.
Our results show a large reduction of the theoretical
uncertainties in predictions of the forward-backward asymmetry,
and allow a precision determination of the top quark electroweak
couplings at future $e^+e^-$ colliders.
\end{abstract}

\maketitle

\noindent \textbf{Introduction.}
Top-antitop quark pairs can be copiously produced at future International
Linear Collider~(ILC), facilitating a detailed study of top quark
properties~\cite{ILC}. The clean enviorement of lepton collider allows
measurement of the process $e^+e^-\to t\bar{t}$  to very high
accuracy, which also demands high precision in theoretical calculation,
in particular the inclusion of higher order QCD radiative
corrections. In the past, significant theoretical efforts have been
focused on $t\bar{t}$ production at threshold, for which
Next-to-Next-to-Leading Order~(NNLO) QCD corrections are known for
more than a decade~\cite{threshold}, and even next-to-next-to-next-to-leading order QCD
corrections will be available in the near future~\cite{nnnlo}. However, for
$t\bar{t}$ production in the continuum only total cross sections are
known in high energy expansion~\cite{highenergy}. Ingredients for a
fully differential NNLO calculation in the continuum have been obtained by
different groups~\cite{formfactor,realvirtual}. Recently, we
reported a fully differential NNLO calculation for the photon mediated
contributions~\cite{Gao:2014nva}, using a NNLO generalization of
phase space slicing
method~\cite{1210.2808,vonManteuffel:2014mva}.~\footnote{Independently,
calculation of inclusive cross section for $e^+e^- \to \gamma^* \to t\bar{t}$ at NNLO based on massive
generalization of antenna subtraction method~\cite{antenna} has been reported recently in ref.~\cite{Dekkers:2014hna}.} In this Letter, we complete this calculation by
including also the SM $Z$ boson contributions.

As an important application of our results, we consider the calculation
of top-quark forward-backward asymmetry~($A_{FB}$) at NNLO in $e^+e^-$
collision in the continuum. In the limit of small top quark mass,
this observable has been computed to NNLO in refs.~\cite{massless}.
The full mass effects are only known for the pure two-loop virtual contributions~\cite{Bernreuther:2006vp}.
We report in this Letter the first calculation of full NNLO QCD corrections to this observable,
including both loop contributions and real-radiation contributions. $A_{FB}$ is an important precision
observable for the determination of neutral-current electroweak couplings of
top quark with photon and $Z$ boson. Their precise measurement is
an important probe of physics beyond Standard Model, {\it e.g.}
Randall-Sundrum models~\cite{Agashe:2005vg},
models of compositeness~\cite{Chivukula:1992ap}. The information of
top quark neutral coupling is encoded in the top-quark form
factor. For on-shell $t\bar{t}$ pair, the form factor can be expressed
by four independent scalar form factors,
\begin{align}
\Gamma^{ttV}_\mu (Q_\mu) &= - i e \left[ \gamma_\mu \Big( F^V_{1v}(Q^2)
+ \gamma_5 F^V_{1a}(Q^2)\Big) 
\right.
\nn
\\
&
\left.
+\left( \frac{\sigma_{\mu\nu}}{2m_t} Q^\nu \Big(i
  F^V_{2v} (Q^2) + \gamma_5 F^V_{2a} (Q^2) \Big)  \right) \right]    
\end{align}
where $Q_\mu$ is the total four-momentum of $t\bar{t}$, $e$ is the
positron charge, and $m_t$ is the top-quark mass. $V$ denotes
photon~($\gamma$) or $Z$ boson. To Leading Order~(LO) in electroweak
theory and QCD, the vector and axial form factors, $F^V_{1v}(Q^2)$ and
$F^V_{1a}(Q^2)$, are given respectively by
\begin{gather}
  F^\gamma_{1v} = Q_t \ , \quad F^\gamma_{1a} = 0 \ ,
\nn
\\
  F^Z_{1v} = \frac{1}{\sin 2\vartheta} \left( \frac{1}{2} - 2 Q_t
    \sin^2 \vartheta \right) \ , \quad F^Z_{1a}  =  \frac{ - 1}{2 \sin
    2\vartheta} 
\end{gather}
$Q_t=2/3$ is the top-quark charge in unit of $e$, and
$\vartheta$ is the weak-mixing angle.
At ILC the top-quark forward-backward asymmetry can be measured
to a precision of about one percent in relative, through both
the full hadronic or semi-leptonic channels~\cite{1005.1756,Amjad:2013tlv,Amjad:2014fwa}.
Correspondingly the above form factors will be determined much more precisely as
compared to at the LHC~\cite{Amjad:2013tlv}, and thus provides strong sensitivity
to any new physics that could modify the top-quark electroweak couplings.
In this Letter, we computed the NNLO QCD corrections to the fully differential
production of top quark pair, thereby obtain the $A_{FB}$ at NNLO for the first
time. Our calculation provides the most precise predictions on $A_{FB}$ including
its theoretical uncertainties, and also allows corrections for experimental
acceptance using the full kinematic information.

%The differential cross section
%at LO for the
%production of $t\bar{t}$ pair with polar angle $\theta_t$ can be written
%as~\cite{PHRVA.D25.1218}
%\begin{align}
%  \frac{\dd \sigma}{\dd \cos\theta_t} = \frac{3}{8} \sigma_U (1+\cos^2\theta_t)
%   + \frac{3}{4}\sigma_L \sin^2\theta_t  + \frac{3}{4}
%  \sigma_F \cos\theta_t 
%\label{eq:cos}
%\end{align}
%where the unpolarized transverse cross section $\sigma_U$,
%longitudinally polarized cross section $\sigma_L$, and difference
%between left and right polarization cross section $\sigma_F$ can be
%found in ref.~\cite{PHRVA.D25.1218}. It is clear that only $\sigma_F$, which is due
%to the interference of vector and axial current, contributes to
%$A_{FB}$. In this Letter, we computed the NNLO QCD corrections to
%the fully differential production of top quark pair,
%thereby obtain the $A_{FB}$ at NNLO for the first time.

\noindent \textbf{The formalism.} A fully differential calculation for
$e^+e^- \to t\bar{t}$ at NNLO in QCD involves three types of
diagrams, namely the two-loop virtual diagrams, one-loop real-virtual diagrams,
and double real-emission diagrams. The individual contributions of these diagrams
are known for some time, but combining them in a consistent way is a
non-trivial task due to the presence of infrared singularities in QCD matrix elements. To this end, we employ a NNLO generalization phase-space
slicing method, described in detail in a previous
publication~\cite{Gao:2014nva}. We briefly summarize its essential features here.

In perturbative QCD, differential cross section for any infrared-safe
observable $O$ has the schematic form
\begin{align}
  \frac{\dd \sigma}{\dd O} = \int\! \dd PS_{t\bar{t}+X}
  |\mathcal{M}_{e^+e^-\to t\bar{t} X}|^2 \delta\Big(O -
  F(\{p_i\})\Big) \ ,
\label{eq:full}
\end{align}
where $O$ is calculated from the set of final-state momentum $\{p_i\}$ through
the measurement function $F$. Inserting a unit decomposition $1 =
\Theta(\lambda - E_X) + \Theta( E_X - \lambda) \equiv \Theta_\mathrm{I} + \Theta_{\mathrm{II}}$, we can
write Eq.~(\ref{eq:full}) as
\begin{align}
  \frac{\dd \sigma}{\dd O} = \frac{\dd \sigma_\mathrm{I}}{\dd O} + \frac{\dd
    \sigma_{\mathrm{II}}}{\dd O} \ ,
\end{align}
where 
\begin{align}
  \frac{\dd \sigma_\mathrm{I}}{\dd O} = & \int\! \dd PS_{t\bar{t}+X}
  |\mathcal{M}_{e^+e^-\to t\bar{t} X}|^2 \delta\Big(O -
  F(\{p_i\})\Big) \Theta_\mathrm{I}\ ,
\nn
\\
  \frac{\dd \sigma_{\mathrm{II}}}{\dd O} = & \int\! \dd PS_{t\bar{t}+X}
  |\mathcal{M}_{e^+e^-\to t\bar{t} X}|^2 \delta\Big(O -
  F(\{p_i\})\Big) \Theta_{\mathrm{II}}  \nn
\end{align}

Obtaining $\mathcal{O}(\als^2)$ corrections to $\dd \sigma/\dd O$
simply amounts to achieving the same order of accuracy for $\dd \sigma_\mathrm{I}/\dd O$ and
$\dd \sigma_{\mathrm{II}}/\dd O$. For $\dd \sigma_{\mathrm{II}}/\dd O$ this is
simple. The presence of theta function $\Theta_{\mathrm{II}}$ implies that
there is at least one parton other than the $t\bar{t}$ pair in the
final state. This parton has the effect of regulating the so-called
double-unresolved divergences in the QCD matrix elements. The only
infrared divergences~(soft or collinear) are of NLO origin and can be
easily dealt with using any NLO subtraction scheme~\cite{subtraction1,subtraction2}. In other words,
the $\mathcal{O}(\als^2)$ contributions to $\dd \sigma_{\mathrm{II}}/\dd O$ can be obtained from a standard NLO QCD
calculation for $e^+e^- \to t\bar{t} j$~\cite{realvirtual}. The results will
exhibit logarithmic singular dependence on the artificial parameter
$\lambda$. Be specific, we employ the massive version
of dipole subtraction method~\cite{subtraction2}. The one-loop
real-virtual calculation is carried out by the automated
program \texttt{GoSam2.0}~\cite{Cullen:2014yla} with loop integral reductions from
\texttt{Ninja}~\cite{Mastrolia:2012bu,Peraro:2014cba} and scalar integrals from \texttt{OneLOop}~\cite{vanHameren:2009dr,vanHameren:2010cp}.
Note that when $\sqrt{s}>4m_t$, the
channel for production of $t\bar{t} t\bar{t}$ is open. However,
these additional contributions are themselves infrared finite and
small for the energy range considered here, thus are not included.
Similarly, we do not include the real emmision diagrams of which
$\gamma^*/Z^*$ couples to light or bottom quarks and the top quarks
emitting from gluon splliting. Those contributions are also small,
and espeially do not contribute to the top-quark FB asymmetry. 

The calculation of $\dd \sigma_{\mathrm{I}}/\dd O$ is substantially
more involved. However, significant simplification can be achieved if $\lambda \ll
m_t$~\footnote{We count $m_t$ and $\sqrt{s}$, the center-of-mass energy, the
same order.}. In that regime, universal factorization properties of
QCD matrix elements allow to write the distribution as 
\emph{soft-virtual} contributions plus power suppressed terms
\begin{align}
 \frac{ \dd \sigma_{\mathrm{I}}}{\dd O} = \frac{\dd \sigma_{\mathrm{s.v.}}}{\dd O} +
  \mathcal{O}\left(\frac{\lambda}{m_t} \right) \ ,
\end{align}
where a soft expansion has been performed in $\dd
\sigma_{\mathrm{s.v.}}/\dd O$ through the phase-space volume, the matrix
elements, and the measurement function. 
As explained in ref.~\cite{Gao:2014nva}, the soft-virtual contributions have
the form of factorized product of hard function and soft function,
each of which is known to $\mathcal{O}(\als^2)$ in analytical form. The hard function is
essentially the heavy quark form factors calculated in refs.~\cite{formfactor}, and the soft
function is the phase-space integral in the
eikonal limit. The soft function is
the same for $\gamma$ or $Z$ meadiated contributions. Comparing with
the vector contributions calculated in ref.~\cite{Gao:2014nva}, the only
difference is the inclusion of axial-vector and anomaly contributions
in the hard function. We note that $\dd \sigma_{\mathrm{s.v.}}/\dd O$
also exhibits logarithmic singular dependence on $\lambda$. 

Combining $\dd \sigma_{\mathrm{s.v.}}/\dd O$ and $\dd \sigma_{\mathrm{II}}/\dd
O$, we obtain a formally exact results for $\dd \sigma/\dd O$,
in the limit $\lambda \to 0$. However, such a limit can never be reached
because $\dd \sigma_{\mathrm{II}}/\dd O$ is usually computed
numerically. In pratice, we choose the parameter sufficently small
such that the power suppressed terms can be safely neglected, and the
kinematical approximation in the soft-virtual contributions can be
justified. The appropriate choice of $\lambda$ can be indicated
by searching for an region in which the dependence of $\lambda$ in  $\dd \sigma_{\mathrm{s.v.}}/\dd O +
\dd \sigma_{\mathrm{II}}/\dd O$ is minimized.

\noindent \textbf{Total cross sections.}
We first present our numeric results on total cross sections. We use two-loop running of the QCD coupling
constants with $N_l=5$ active quark flavors
and $\alpha_s(M_Z)=0.118$. We choose the $G_F$ parametrization scheme~\cite{Denner:1990ns} for the electroweak
couplings with $M_W=80.385\,{\rm GeV}$, $M_Z=91.1876\,{\rm GeV}$, $M_{t}=173\,{\rm GeV}$,
and $G_F=1.166379\times 10^{-5}\,{\rm GeV}^{-2}$~\cite{Beringer:1900zz}. The renormalization scale is
set to the center of mass energy $\sqrt s$ unless otherwise specified.
The production cross sections through to NNLO in QCD can be expressed as
\begin{equation}\label{eq:kfa}
\sigma_{NNLO}=\sigma_{LO}\left(1+\Delta^{(1)}+\Delta^{(2)}\right),
\end{equation}
where $\Delta^{(1)}$~($\Delta^{(2)}$) denotes the $\ord{\als}$~($\ord{\als^2}$) QCD corrections. Analytical results for $\Delta^{(2)}$ are presented
for production near threshold~\cite{threshold} or by
high energy expansions~\cite{highenergy}
with which we compare our numerical results.

\begin{figure}[!h]
  \begin{center}
  \includegraphics[width=0.4\textwidth]{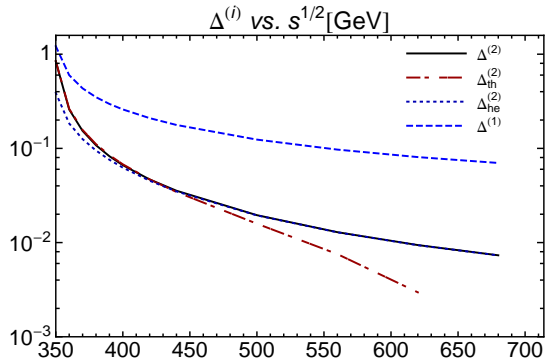}
  \end{center}
  \vspace{-1ex}
  \caption{\label{fig:scan1}
  Comparison of $\ordb$ corrections to total cross sections, $\Delta^{(2)}$, with
  the threshold results $\Delta^{(2)}_{th}$
  and high-energy expansion results $\Delta^{(2)}_{he}$, as functions of
  collision energies.}
\end{figure}

Fig.~\ref{fig:scan1} shows detailed comparison of our numerical results
with the threshold~\cite{threshold} and
high-energy expansion results~\cite{highenergy} in a
wide range of collision energies. It can be seen that our full results works
well in the entire energy region,
i.e., approaching the threshold results at lower energies and the high-energy
expansions on the other end. The $\ord{\als^2}$
correction can reach as large as 80\% for $\sqrt s=350\,{\rm GeV}$, due to threshold Coulomb singularities. On the other hand it is
about 2\% for intermediate
collision energies and deacrease quickly to below 1\% for high energies.
The good agreements of our results on total cross sections with ones from
threshold and high-energy expansions in the corresponding energy region
furhther demonstrate the validity of our calculation.  

\begin{figure}[!h]
  \begin{center}
  \includegraphics[width=0.4\textwidth]{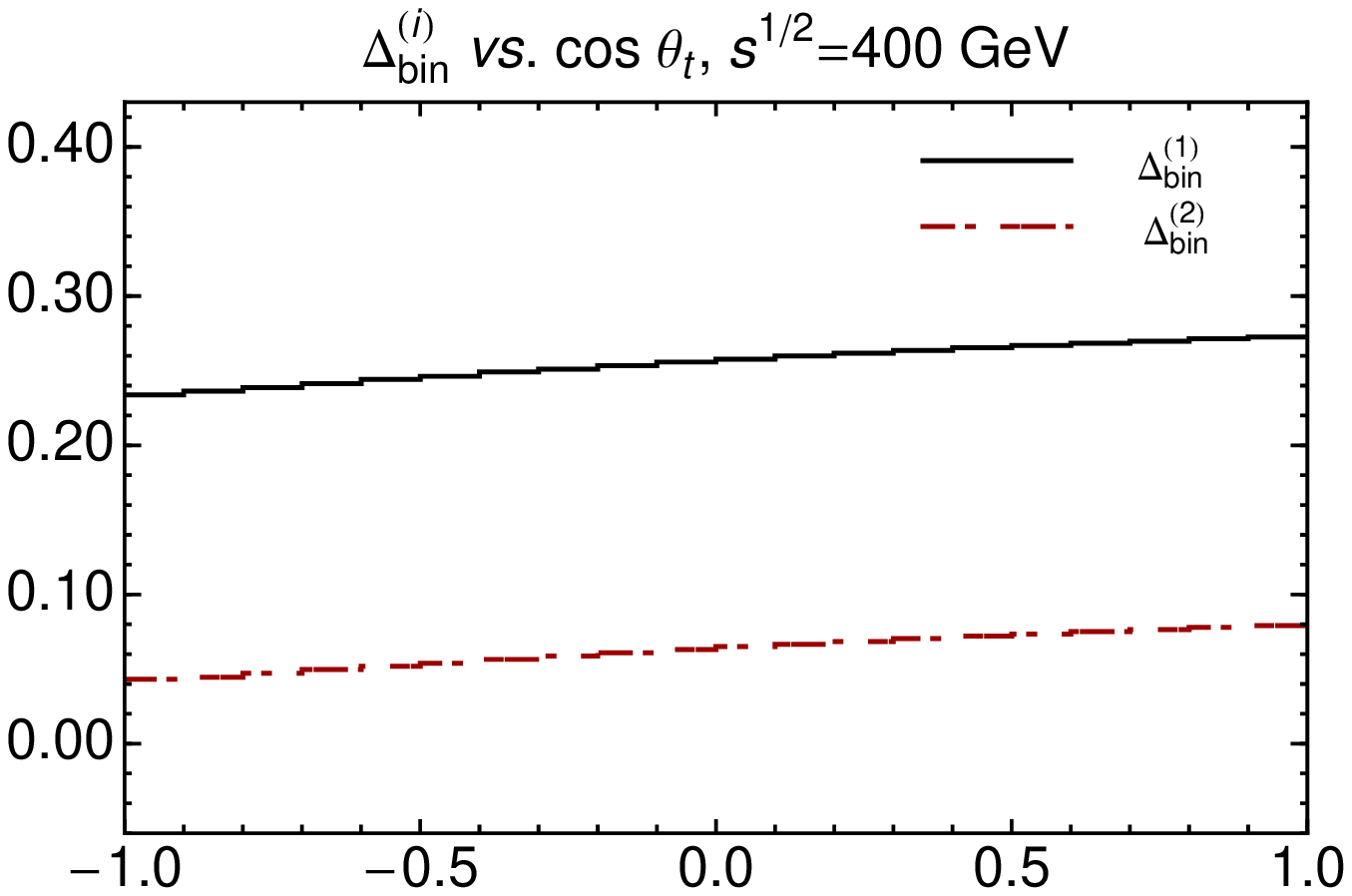}
  \end{center}
  \vspace{-1ex}
  \caption{\label{fig:scan2}
  $\orda$ and $\ordb$ corrections in different $\cos\theta$ bins of top quark,
  $\Delta^{(1)}_{bin}$ and $\Delta^{(2)}_{bin}$, for $\sqrt s=400$ GeV.}
\end{figure}

\begin{figure}[!h]
  \begin{center}
  \includegraphics[width=0.4\textwidth]{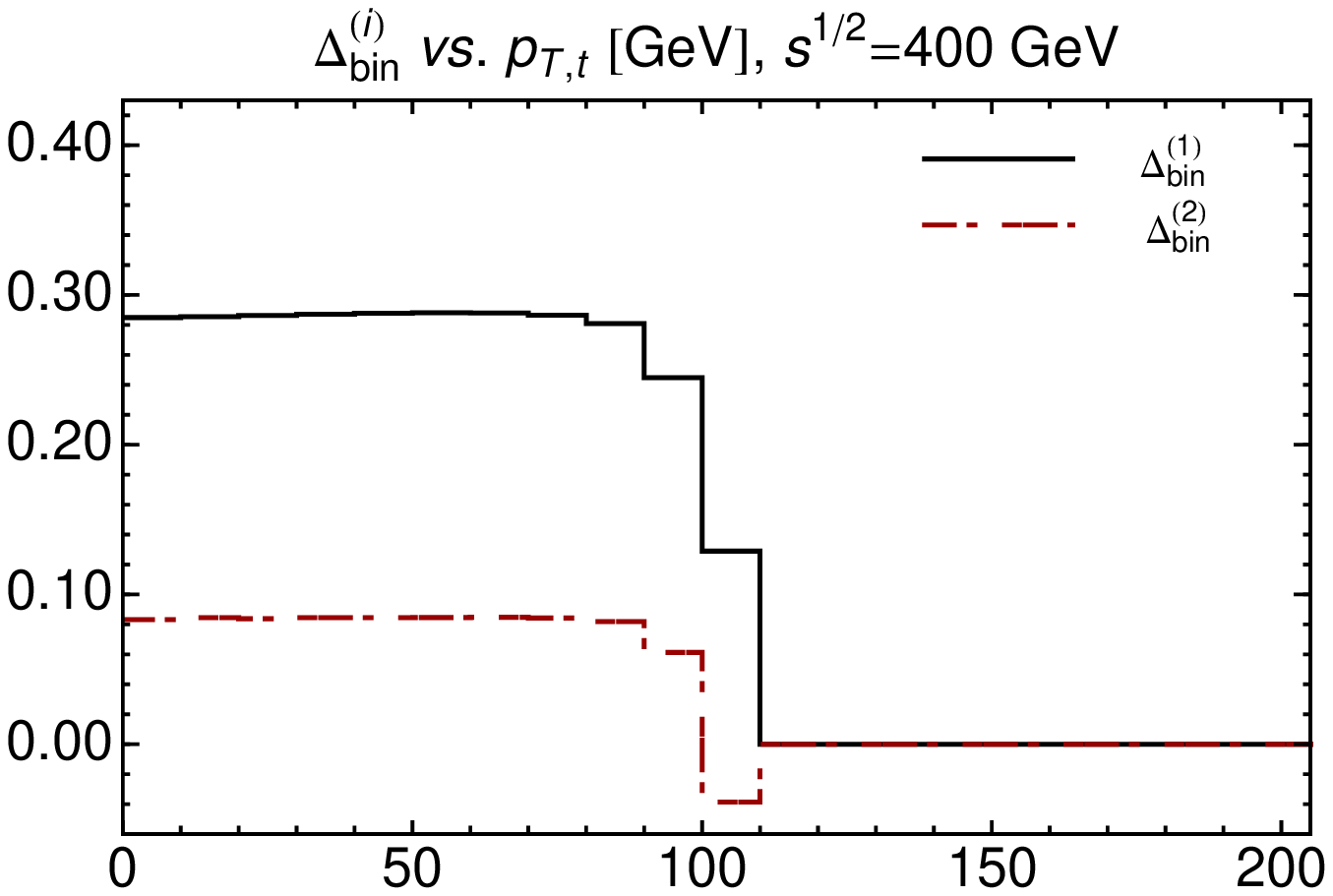}
  \end{center}
  \vspace{-1ex}
  \caption{\label{fig:scan3}
  $\orda$ and $\ordb$ corrections in different $p_T$ bins of top quark,
  $\Delta^{(1)}_{bin}$ and $\Delta^{(2)}_{bin}$, for $\sqrt s=400$ GeV.}
\end{figure}

\noindent \textbf{Differential distributions and $A_{FB}$.}
We can calculate fully differential distributions up to NNLO in QCD based
on the phase space slicing method. At LO, there is only one non-trivial kinematic
variable, which we can choose either as cosine of the scattering angle between
the final-state top quark and the initial-state electron $\cos\theta_t$, 
or transverse momentum of the top quark with respect to the beam line direction $p_{T,t}$.
Similar to the inclusive cross section, we can define the $\orda$ and $\ordb$ corrections
for each kinematic bin, $\Delta^{(1)}_{bin}$ and $\Delta^{(2)}_{bin}$, in analogy to Eq.~(\ref{eq:kfa}).
The results are shown in Fig.~\ref{fig:scan2} for $\cos\theta_t$ and Fig.~\ref{fig:scan3}
for $p_{T,t}$ distributions with a typical collision energies of 400 GeV. The $\ordb$
corrections are about one fourth of the $\orda$ corrections for the total cross section.
However, they show a different kinematic dependence. From Fig.~\ref{fig:scan2} we can
see both the $\orda$ and $\ordb$ corrections are larger in forward direction and thus will
increase the FB asymmetry. Moreover, the differences of $\Delta^{(2)}_{bin}$ in forward and
backward region are of similar size as for $\Delta^{(1)}_{bin}$. Thus the $\ordb$ corrections
to $A_{FB}$ are as important as the $\orda$ corrections as will be shown later.
The transverse momentum distribution in Fig.~\ref{fig:scan3} shows a different feature
comparing to the angular distribution since they are also affected by the energy spectrum
of the top quark. The real corrections pull the energy spectrum to the lower end and
thus the $p_{T,t}$ distribution as well. As shown in Fig.~\ref{fig:scan3}, the $\ordb$
corrections start as positive in low $p_T$ and then decrease to negative
values near the kinematic limits. The $\ordb$ corrections show a relatively larger
impact in the $p_{T,t}$ distribution.

\begin{figure}[!h]
  \begin{center}
  \includegraphics[width=0.4\textwidth]{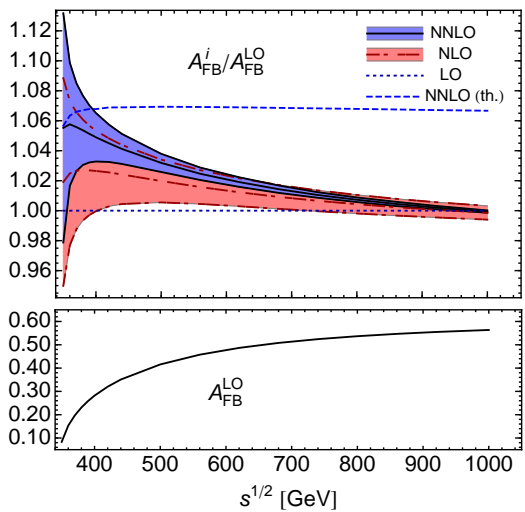}
  \end{center}
  \vspace{-1ex}
  \caption{\label{fig:scan4}
   Lower inset: top quark $A_{FB}$ at the LO as a function of collision energy;
   upper inset: ratios of $A_{FB}$ at the NLO and NNLO to $A_{FB}$ at the LO. The threshold approximation is denoted as \texttt{th.}.}
\end{figure}

The forward-backward asymmetry $A_{FB}$ is defined as the number of top quark
observed in the forward direction minus the one observed in the backward
direction, divided by the total number of top quark observed,
\begin{equation}\label{eq:afb}
A_{FB}=\frac{\sigma_A}{\sigma_S}\equiv
\frac{\sigma(\cos\theta_t>0)-\sigma(\cos\theta_t<0)}
{\sigma(\cos\theta_t>0)+\sigma(\cos\theta_t<0)},
\end{equation}
%with $\theta_t$ be the angle between the top-quark outgoing direction and the electron
%incidnet direction.
We show $A_{FB}$ at LO as a function of the collision energy in
the lower inset of Fig.~\ref{fig:scan4}. The $A_{FB}$ at NLO and NNLO are calculated by
using the corresponding NLO and NNLO cross sections in both the denominator and numerator
of Eq.~(\ref{eq:afb}), and are shown in the upper inset of
Fig.~\ref{fig:scan4} normalized to the $A_{FB}$ at LO. The $\orda$ correction is about
2\% for $\sqrt s$ around 500 GeV. The $\ordb$ correction further increases $A_{FB}$ by
about 1.2\% in the same region. We also plot the $A_{FB}$ calculated
using the two-loop threshold cross sections~\cite{Bernreuther:2006vp} for comparison,
which shows good agreement with our exact result in energy region just above the production
threshold. This is expected since in the threshold region the former ones are dominant.   
We further investigate uncertainties of predictions on $A_{FB}$ due to missing corrections
beyond NNLO. A conventional way to estimate those uncertainties is by checking the residual
QCD scale dependence. However, for ratios like $A_{FB}$, if we vary the scales simultaneously
in $\sigma_A$ and $\sigma_S$, it tends to be too optimistic. For example, the NLO prediction
with scale uncertainty does not overlap with the NNLO prediction. Thus a more appropriate
prescription is to vary the scales independently in $\sigma_A$ and $\sigma_S$. We change
the scale by a factor of two upward and downward in both $\sigma_A$ and $\sigma_S$, and
add the fractional uncertainties to $A_{FB}$ in quadrature. The resulting uncertainties
are shown in Fig.~\ref{fig:scan4} by the colored bands. With the $\ordb$ corrections
the scale uncertainty on $A_{FB}$ has been reduced to less than half of the value at NLO
as further shown in Table.~\ref{tab:scan1}. The third and fourth columns of Table.~\ref{tab:scan1}
show the NLO and NNLO predictions of $A_{FB}$ together with the scale uncertainties all
shown in percentage. The column $\delta A_{FB}^{NNLO}$ represents variation of FB
asymmetry due to uncertainty of top-quark mass input, which is taken to be $\pm 0.5$ GeV
simply for comparison. For a typical collision energy of 500 GeV, the residual scale
uncertainty of NNLO prediction on $A_{FB}$ is 0.0025 or half percent in relative,
which is well below the projected experimental precision of future ILC~\cite{1005.1756}.
The uncertainty due to top quark mass input is relatively small especially considering
the projected precision on mass measurement at the ILC.

\begin{table}

\begin{tabular}{|c|c|c|c|c|}
\hline
$\sqrt s$ [GeV] & $A_{FB}^{LO}$ & $A_{FB}^{NLO}$ & $A_{FB}^{NNLO}$ 
& $\delta A_{FB}^{NNLO}$ \tabularnewline
\hline
400 & 28.20 & $28.94\pm0.76$ & $29.58\pm0.46$ & $\pm 0.26$ \tabularnewline
\hline
500 & 41.56 & $42.39\pm0.59$ & $42.89\pm0.25$ & $\pm 0.12$ \tabularnewline
\hline
800 & 53.68 & $53.91\pm0.33$ & $54.07\pm0.08$ & $\pm 0.04$ \tabularnewline
\hline
\end{tabular}

\caption{Top-quark forward-backward asymmetry at different perturbative orders
for representative $\sqrt s$ values. All
values are shown in percentage.
    \label{tab:scan1}}
\end{table}

\begin{figure}[!h]
  \begin{center}
  \includegraphics[width=0.4\textwidth]{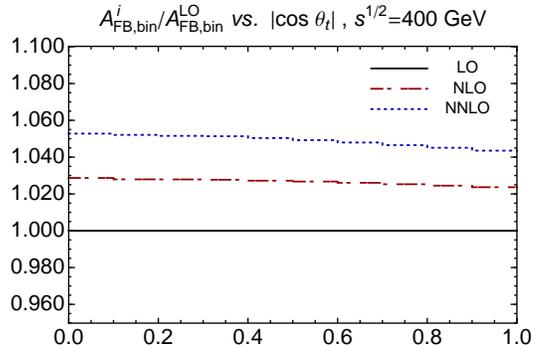}
  \end{center}
  \vspace{-1ex}
  \caption{\label{fig:scan5}
   Top-quark forward-backward asymmetry in different $|\cos\theta_t|$ bins, $A_{FB,bin}$,
   normalized to the LO predictions, for $\sqrt s=400$ GeV.}
\end{figure}

We can also look at top-quark FB asymmetry at a more exclusie level, namely the
FB asymmetry in different $|\cos\theta_t|$ bins, $A_{FB,bin}$. In Fig.~\ref{fig:scan5}
we plot ratios of the NLO and NNLO predictions on $A_{FB,bin}$ to the LO ones for
$\sqrt s=400$ GeV. Both the $\orda$ and $\ordb$ corrections are almost flat on
$|\cos\theta_t|$ for $\sqrt s =400$ GeV, but decrease slightly with $|\cos\theta_t|$
for $\sqrt s=500$ GeV which is not shown here due to limited space. 

\noindent \textbf{Conclusions.} We have presented the first complete NNLO QCD corrections to
top-quark pair production at $e^+e^-$ collisions. The calculation is at fully differential
level based on a generalization of phase slicing method to NNLO in QCD~\cite{Gao:2014nva}. Especially we study
in detail the corrections to top-quark forward-backward asymmetry $A_{FB}$. The NNLO correction to
$A_{FB}$ is half of the size of the NLO corrections or even larger, for a typical collision
energy of $400\sim 500$ GeV at future linear colliders. Moreover, our results show a large
reduction in the theoretical uncertainties on predictions of $A_{FB}$. The residual scale uncertainty
is well below the projected experimental precision. Our results allow a
precise determination of the top-quark neutral-current couplings at future linear
colliders and thus the probe of various new physics beyond the SM.
Besides, there could be several interesting applications of the method and results presented here. Firstly, it would be interesting to
apply this calculation to charm and bottom quark production at Z boson
mass pole. 
Secondly, decay of Higgs boson to massive quark can be calculated in the same way to NNLO in QCD, since the two-loop matrix elements are available~\cite{Bernreuther:2005gw}.
Thirdly, it should be straightforward to combine production and
leptonic decay~\cite{1210.2808,Brucherseifer:2013iv} of top-quark pair in $e^+e^-$ collisions within narrow width
approximation at NNLO. Last but not least, our calculation may also be used to improve the accuracy of
event shape resummation related to heavy quark mass
measurement~\cite{Fleming:2007qr}. 

\begin{acknowledgments}
This work was supported by the U.S. DOE under contract \textrm{DE-AC02-76SF00515}, Early Career Research Award
\textrm{DE-SC0003870} by Lightner-Sams Foundation, and Munich Institute for Astro- and Particle Physics (MIAPP) of the DFG cluster of excellence "Origin and Structure of the Universe". Work at ANL is supported in part by the U.S. Department of Energy under Contract No. \textrm{DE-AC02-06CH11357}. HXZ thanks the theory group of Paul Scherrer Institute~(PSI) at Zurich and the Center for Future High Energy Physics~(CFHEP) at Beijing for hospitality where the project is finalized.
\end{acknowledgments}

\end{document}